\documentclass[preprint2]{aastex}

\usepackage{graphicx}
\usepackage{url}
\usepackage{natbib}

\newcommand{\ie}{{\it i.e.}}
\newcommand{\eg}{{\it e.g.}}
\newcommand{\nobs}{n_{\rm obs}}
\newcommand{\nvar}{n_{\rm var}}
\newcommand{\nvec}{n_{\rm vec}}

\newcommand{\X}{\mathbf{X}}
\newcommand{\PP}{\mathbf{P}}   
\newcommand{\C}{\mathbf{C}}
\newcommand{\W}{\mathbf{W}}
\newcommand{\V}{\mathbf{V}}

\newcommand{\Xr}{\vec \X^\mathrm{row}}
\newcommand{\Xc}{\vec \X^\mathrm{col}}
\newcommand{\Cr}{\vec \C^\mathrm{row}}
\newcommand{\Cc}{\vec \C^\mathrm{col}}

\newcommand{\PPc}{\vec \PP^\mathrm{col}}
\newcommand{\Wr}{\vec \W^\mathrm{row}}
\newcommand{\Wc}{\vec \W^\mathrm{col}}

\shorttitle{PCA with Noisy & Missing Data}
\shortauthors{S. Bailey}

\begin{document}

\author{Stephen Bailey}
\affil{Lawrence Berkeley National Lab, 1 Cyclotron Rd, Berkeley, CA 94720, USA}

\title{Principal Component Analysis with \\ Noisy and/or Missing Data}

\begin{abstract}
We present a method for performing Principal Component Analysis (PCA) on
noisy datasets with missing values.
Estimates of the measurement error are used to weight the input data
such that compared to classic PCA, the resulting eigenvectors
are more sensitive to the true underlying signal variations rather than
being pulled by heteroskedastic measurement noise.
Missing data is simply the limiting case of weight=0.
The underlying algorithm is a noise weighted Expectation Maximization (EM)
PCA, which has additional benefits of implementation speed and flexibility
for smoothing eigenvectors to reduce the noise contribution.
We present applications of this method on simulated data
and QSO spectra from the Sloan Digital Sky Survey.
\end{abstract}


\section{Introduction}

Principal Component Analysis (PCA) is a powerful and widely used
technique to analyze data by forming a custom
set of ``principal component'' eigenvectors
that are optimized to describe the most
data variance with the fewest number of components
\citep{Pearson1901, Hotelling1933, Jolliffe2002}.
With the full set of eigenvectors the data may be reproduced exactly,
\ie, PCA is a transformation which can lend insight by identifying
which variations in a complex dataset are most significant
and how they are correlated.
Alternately, since the eigenvectors are optimized and sorted by
their ability to describe variance in the data, PCA may be used
to simplify a complex dataset into a few eigenvectors plus coefficients,
under the approximation that higher-order eigenvectors are predominantly
describing fine tuned noise or otherwise less important features of
the data.
Example applications within astronomy include classifying spectra by
fitting them to PCA templates \citep{Paris2011, ConnollySzalay1999},
describing Hubble Space Telescope point spread function variations
\citep{Jee2007},
and reducing the dimensionality of Cosmic Microwave Background map data
prior to analysis \citep{Bond1995}.

A limitation of classic PCA is that it does not distinguish between
variance due to measurement noise {\it vs}. variance due to genuine
underlying signal variations.
Even when an estimate of the
measurement variance is available, this information is not used when
constructing the eigenvectors, {\it e.g.}, by deweighting noisy data.

A second limitation of classic PCA is the case of missing data.
In some applications, certain observations may be missing some
variables and the standard formulas for constructing the
eigenvectors do not apply.  For example within astronomy,
observed spectra do not cover the same rest-frame wavelengths
of objects at different redshifts, and some wavelength bins may
be masked due to bright sky lines or cosmic ray contamination.
Missing data is an extreme case of noisy data,
where missing data is equivalent to data with infinite measurement variance.

This work describes a PCA framework which incorporates
estimates of measurement variance while solving for the principal components.
This optimizes the eigenvectors to describe the true underlying signal
variations without being unduly affected by known measurement noise.
Code which implements this algorithm is available at
\url{https://github.com/sbailey/empca}~.

\citet{Jolliffe2002} \S 13.6 and \S 14.2
review prior work on PCA with missing data and
incorporating weights into PCA.
Most prior work focuses on the identification and removal of outlier data, 
interpolation over missing data,
or special cases such as when the weights can be factorized
into independent per-observation and per-variable weights.
\citet{GabrielZamir1979}, \citet{Wentzell1997}, \citet{TippingBishop1999},
and \citet{SrebroJaakkola2003}
present iterative solutions for the case of general weights, though
none of these find the true PCA solution with orthogonal eigenvectors 
optimally ranked by their ability to describe the data variance.
Instead, they find an unsorted set of non-orthogonal vectors which
as a set are optimized to describe the data variance, but individually
they are not the optimal linear combinations to describe the most
variance with the fewest vectors.
Their methods are sufficient for weighted lower-rank matrix
approximation, but they lack the potential data insight from optimally
combining and sorting the eigenvectors to see which components contribute
the most variation.

Within the astronomy literature,
\citet{ConnollySzalay1999} discuss how to interpolate over missing data and
use PCA eigenspectra to \emph{fit}
noisy and/or missing data, but they do not address the case of how to
\emph{generate} eigenspectra from noisy but non-missing data.
\citet{BlantonRoweis2007} generate template spectra from noisy and
missing data using Non-negative Matrix Factorization (NMF).
This method is similar to PCA with the constraint that the template
spectra are strictly positive, while not requiring the templates to
be orthonormal.  \citet{TsHogg2012} present a more general
``Heteroskedastic Matrix Factorization'' approach to study
Sloan Digital Sky Survey spectra while properly accounting for
measurement noise.
Their underlying goal is similar to this work,
though with an algorithmically different implementation.

The methods presented here directly solve for the PCA eigenvectors
with an iterative solution based upon Expectation Maximization PCA (EMPCA).
\citet{Roweis1997} describes an unweighted version of EMPCA, including a method
for interpolating missing data, but he does not address the issue
of deweighting noisy data.
We also take advantage of the iterative nature of the solution to bring
unique extensions to PCA, such as noise-filtering the eigenvectors
during the solution.

The approach taken here is fundamentally pragmatic.  For example, if
one is interested in generating eigenvectors to describe 99\% of the
signal variance, it likely doesn't matter if an iterative algorithm
has ``only'' converged at the level of $10^{-5}$ even if the machine
precision is formally $10^{-15}$.  We discuss some of the limitations
of Weighted EMPCA in \S\ref{sec:discussion}, but ultimately we find that
these issues are not limiting factors for practical applications.

This work was originally developed for PCA analysis of
astronomical spectra and examples are given in that context.
It should be noted, however, that these methods are
generally applicable to any PCA application with noisy and/or
missing data --- nothing in the underlying methodology is specific to
astronomical spectra.

\section{Notation}
\label{sec:notation}

This paper uses the following notation:
Vectors use arrows, $\vec x$, while $x_i$ represents the scalar
element $i$ of vector $\vec x$.
For sets of vectors, $\vec x_j$ represent vector number $j$
(not element $j$ of vector $\vec x$).
Matrices are in boldface, $\X$.  To denote vectors formed by
selected columns or rows of a matrix, we use
$\Xc_j$ for the vector formed from column $j$ of matrix $\X$ and
$\Xr_i$ for the vector formed from row $i$ of matrix $\X$.
The scalar element at row $i$ column $j$ of matrix $\X$ is
$\X_{ij}$.

For reference, we summarize the names of the primary variables here:
$\X$ is the data matrix with $\nvar$ rows of variables
and $\nobs$ columns of observations.
$\PP$ is the PCA eigenvector matrix with $\nvar$ rows of variables
and $\nvec$ columns of eigenvectors; $\vec\phi$ is a single eigenvector.
These eigenvectors may fit
the data using a matrix of coefficients $\C$, where
$\C_{kj}$ is the contribution of eigenvector $k$ to observation $j$.
Indices $i$, $j$, and $k$ index variables, observations, and eigenvectors
respectively.  
$\vec\X$ is the vector formed by stacking the columns of matrix $\X$.
The \emph{measurement} covariance of dataset $\vec\X$ is $\V$, while
$\W$ is the weights matrix for dataset $\X$ for the case of
independent measurement noise such that $\W$ has the same dimensions as $\X$.

\section{Classic PCA}
\label{sec:ClassicPCA}

For an accessible tutorial on classic PCA, see \citet{ShlensPCATutorial}.
A much more complete treatment is found in \cite{Jolliffe2002}.
Algorithmically, the steps are simple: The principal components
$\{\vec \phi_k\}$
of a dataset are simply the eigenvectors of the covariance of that dataset,
sorted by their descending eigenvalues.
A new observation $\vec y$ may be approximated as
\begin{equation}
\label{eq:recon}
\vec y = \vec \mu + \sum_k c_k \vec \phi_k
\end{equation}
where $\vec\mu$ is the mean of the initial dataset and
$c_i$ is the reconstruction coefficient for eigenvector $\vec \phi_i$.
For the rest of this
paper, we will assume that $\vec\mu$ has already been subtracted from
the data, \ie, $\vec y \leftarrow (\vec y - \vec \mu)$.

To find a particular coefficient $c_{k^\prime}$, take
the dot product of both sides with $\vec \phi_{k^\prime}$, noting that because
of the eigenvector orthogonality,
$\vec \phi_k \cdot \vec \phi_{k^\prime} = \delta_{kk^\prime}$ (Kroeneker-delta):
\begin{eqnarray}
\vec y \cdot \vec\phi_{k^\prime}
    & = & \sum_k c_k \vec\phi_k \cdot \vec\phi_{k^\prime}  \label{eq:alphadot1} \\
    & = & \sum_k c_k \delta_{kk^\prime}          \label{eq:alphadot2} \\
    & = & c_{k^\prime}                             \label{eq:alphadotn} 
\end{eqnarray}

Note that the neither the solution of $\{\vec \phi_k\}$
nor $\{c_k\}$ makes use of any noise estimates or weights for the data.
As such, classic PCA solves the minimization problem
\begin{equation}
    \chi^2 = \sum_{i,j} [\X - \PP \C]_{ij}^2
\end{equation}
where $\X$ is a dataset matrix whose columns are observations
and rows are variables,
$\PP$ is a matrix whose columns are the principal components
$\{\vec\phi_k\}$ to find,
and $\C$ is a matrix of coefficients to fit $\X$ using $\PP$.
For clarity, the dimensions of these matrices are:
$\X[n_\mathrm{var}, n_\mathrm{obs}]$,
$\PP[n_\mathrm{var}, n_\mathrm{vec}]$, and
$\C[n_\mathrm{vec}, n_\mathrm{obs}]$,
where $n_\mathrm{obs}$, $n_\mathrm{var}$, and $n_\mathrm{vec}$ are
the number of observations, variables, and eigenvectors respectively.
For example, when performing PCA on spectra, $n_\mathrm{obs}$ is
the number of spectra, $n_\mathrm{var}$ is the number of wavelength
bins per spectrum, and $n_\mathrm{vec}$ is the number of eigenvectors
used to describe the data, which may be smaller than the total
number of possible eigenvectors.

\section{Adding Weights to PCA}

The goal of this work is to solve for the eigenvectors $\PP$
while incorporating a weights matrix $\W$ on the dataset $\X$:
\begin{equation}
    \chi^2 = \sum_{i,j} \W_{ij}[\X - \PP \C]_{ij}^2
\end{equation}
We also describe the more general cases of per-observation covariances
$\V_j$:
\begin{equation}
    \chi^2 =
    \sum_{\mathrm{obs}~j}
        \left( \Xc_j - \PP \Cc_j \right)^T
        \V_j^{-1}
        \left( \Xc_j - \PP \Cc_j \right)
\end{equation}
where we have used the notation that $\Xc_j$ is the vector formed from the
$j$th column of the matrix $\X$,
and similarly for $\Cc_j$.
In the most general case, there is covariance $\V$ between all
variables of all observations, {\it i.e.}, we seek to minimize
\begin{equation}
    \label{eq:weighted_chi2}
    \chi^2 =
    \left( \vec{\mathbf X} - [\mathbf{P}] \vec\mathbf{C} \right)^T
    \V^{-1}
    \left( \vec{\mathbf X} - [\mathbf{P}] \vec\mathbf{C} \right)
\end{equation}
Where $\vec{\mathbf X}$ and $\vec{\mathbf C}$ are the vectors formed by
concatenating all columns of $\X$ and $\C$, and
$[\mathbf{P}]$ is the matrix formed by stacking $\PP$ $\nobs$ times.

This allows one to incorporate error estimates on heteroskedastic
data such that particularly noisy data does not unduly influence
the solution.  We will solve this problem using an iterative method
known within the statistics community as ``Expectation Maximization''.

\section{Weighted Expectation Maximization PCA}
\label{sec:WEMPCA}

\subsection{Expectation Maximization PCA}

Expectation Maximization (EM) is an iterative technique for solving
parameters to maximize a likelihood function for models with unknown
hidden (or latent) variables \citep{DLR1977}.
Each iteration involves two steps:
finding the expectation value of the hidden variables given the current
model (E-step), and then modifying the model parameters to maximize the fit
likelihood given the estimates of the hidden variables (M-step).

As applied to PCA, the parameters to solve
are the eigenvectors, the latent variables are the coefficients
$\{ c \}$
for fitting the data using those eigenvectors, and the likelihood is
the ability of the eigenvectors to describe the data.
To solve the single most significant eigenvector,
start with a random vector $\vec \phi$ of length $\nvar$.
For each observation $\vec x_j$, solve the coefficient
$c_j = \vec x_j \cdot \vec\phi$
which best fits that observation using $\vec\phi$.
Then using those coefficients, update $\vec\phi$ to find the vector
that best fits the data given those coefficients:
$\vec\phi \leftarrow \sum_j c_j \vec x_j / \sum_j c_j^2$.  Then
normalize $\vec\phi$ to unit length and iterate the solutions to
$\{ c \}$ and $\vec\phi$ until converged.  In summary:
\begin{tabbing}  
\hskip 0.5in \= \hskip 0.25in \= \hskip 0.25in \= \hskip 0.25in \= \hskip 1.0in \= \kill
\+ \kill
$\vec\phi \leftarrow$ random vector of length $\nvar$    \\*
repeat until converged:                               \\*
\>  For each observation $\vec x_j$:                  \\*
\>  \>  $c_j \leftarrow \vec x_j \cdot \vec\phi$ \> \> {\it E-step} \\*
\>  $\vec\phi \leftarrow \sum_j c_j \vec x_j / \sum_j c_j^2$  
    \> \> \> {\it M-step} \\*
\>  $\vec \phi \leftarrow \vec \phi / |\vec\phi|$      \> \> \> \it Renormalize \\*
return $\vec \phi$
\end{tabbing}

\leftskip 0.0in
This generates a vector $\vec\phi$ which is the dominant PCA eigenvector of the
dataset $\X$, where the observations $\vec x_j$ are the columns of $\X$.
The expectation step finds the coefficients $\{c_j\}$ which best fit
$\X$ using $\vec \phi$ (see equations \ref{eq:alphadot1} to \ref{eq:alphadotn}).
The likelihood maximization step then uses those coefficients to update
$\vec\phi$ to minimize
\begin{equation}
    \label{eq:unweighted_chi2}
    \chi^2 = \sum_{\mathrm{var}~i,~\mathrm{obs}~j} ({\bf X}_{ij} - c_j \phi_i)^2
\end{equation}
In practice the normalization $\sum_j c_j^2$ in the M-step is unnecessary
since $\vec\phi$ is renormalized to unit length after every iteration.

At first glance, it can be surprising that this algorithm works at all.
Its primary enabling feature is that at each iteration the coefficients
$\{ c_j \}$ and vector $\vec \phi$ minimize the $\chi^2$ better than
the previous iteration.  
The $\chi^2$ of equation~\ref{eq:unweighted_chi2}
has a single minimum \citep{SrebroJaakkola2003}, thus when any
minimum is found it is the true global minimum.
It is possible, however, to also have saddle points to
which the EMPCA algorithm could converge from particular starting points.
It is easy to test solutions for being at a saddle point and restart the
iterations as needed.

The specific convergence criteria are application specific.  One pragmatic
option is that the eigenvector itself is changing slowly,
\ie, $|\Delta \vec\phi| < \epsilon$.
Alternately, one could require that
the change in likelihood (or $\Delta \chi^2$) from one iteration
to the next is below some threshold.
Convergence and uniqueness will be discussed in $\S$\ref{sec:convergence}
and \S\ref{sec:uniqueness}.  For now we simply note that many PCA applications
are interested in describing 95\% or 99\% of the data variance, and the
above algorithm typically converges very quickly for this level of precision,
even for cases when the formal computational machine convergence may
require many iterations.

To find additional eigenvectors, subtract the projection of $\vec\phi$
from ${\bf X}$ and repeat the algorithm.  Continue this procedure until
enough eigenvectors have been solved that the remaining variance is
consistent with the expected noise of the data, or until enough
eigenvectors exist to approximate the data with the desired fidelity.
If only a few eigenvectors are
needed for a large dataset, this algorithm can be much faster than classic
PCA which requires solving for all eigenvectors whether or not they are
needed.
Scaling performance will be discussed further in $\S$\ref{sec:scaling}.

\subsection{EMPCA with per-observation weights}

The above algorithm treats all data equally when solving for the eigenvectors
and thus is equivalent to classic PCA.
If all data are of approximately equal quality then this is fine, but if some
data have considerably larger measurement noise they can unduly influence
the solution.
In these cases, the high signal-to-noise data should receive greater weight
than low signal-to-noise data.  This is conceptually equivalent to the
difference between a weighted and unweighted mean.

In some applications, it is sufficient to have a single weight per observation
so that all variables within an observation are equally weighted, but different
observations are weighted more or less than others.
In this case, EMPCA can be extended with per-observation weights $w_i$.
The observations $\mathbf{X}$ should have their weighted mean subtracted,
and the likelihood maximization step (M-step) is replaced with:
\begin{equation}
\vec\phi \leftarrow \sum_j w_j c_j \vec x_j
\end{equation}
The normalization denominator has been dropped because we re-normalize
$\vec\phi$ to unit length every iteration.

\subsection{EMPCA with per-variable weights}

If each variable for each observation has a different weight, the situation
becomes more complicated since we cannot use simple dot products to
derive the coefficients $\{c_j\}$.
Instead, one must solve a set of linear equations for $\{c_j\}$.
Similarly, the likelihood maximization step
must solve a set of linear equations to update $\vec \phi$
instead of just performing a simple sum.
The weighted EMPCA algorithm now starts with a \emph{set} of random orthonormal
vectors $\{ \vec \phi_k \}$ and iterates over:
\begin{enumerate}
    \item For each observation $\vec x_j$, solve coefficients $c_{kj}$:
        \begin{equation}        
            \label{eq:expectation}
            \vec x_j = \sum_k c_{kj} \vec \phi_k  
        \end{equation}
    \item Given $\{ c_{kj} \}$, solve each $\vec \phi_k$ one-by-one
        for $k$ in $1..n_{\mathrm{vec}}$:
        \begin{equation}
            \label{eq:maximization}
            \vec x_j - \sum_{k^\prime<k} c_{k^\prime j} \vec\phi_{k^\prime} = c_{kj} \vec\phi_k  \label{eq:phi}
        \end{equation}
\end{enumerate}
Both of the above steps can be solved using weights on $\vec x_j$, thus
achieving the goals of properly weighting the data while solving for the
coefficients $\{ c_{kj} \}$ and eigenvectors $\vec \phi_k$.
Implementation details will be described in the following two sub-sections,
where we will return to using matrix notation.

\subsubsection{Notes on solving $\{c_{kj}\} = \C$}
\label{sec:solve_c}

\begin{figure}[t]
\centering
\plotone{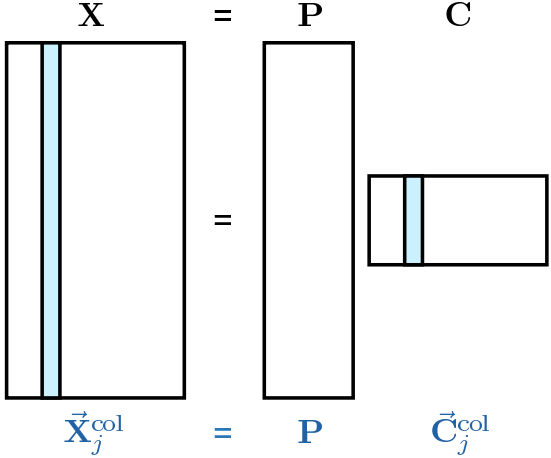}
\caption{Illustration of solving $\C$ one column at a time,
{\it i.e.,} equation \ref{eq:solve_c}.}
\label{fig:solve_c}
\end{figure}

In equation \ref{eq:expectation}, the $\vec\phi_k$ vectors are fixed and one solves
the coefficients $c_{kj}$ with a separate set of equations for each
observation $\vec x_j$.  Written in matrix form, $\X = \PP \C$ can be
solved for each independent observation column $j$ of $\X$ and $\C$:
\begin{equation}
    \label{eq:solve_c}
    \Xc_j = \PP \Cc_j + \mathrm{noise}
\end{equation}
Equation \ref{eq:solve_c} is illustrated in figure~\ref{fig:solve_c}.

Solving equation \ref{eq:solve_c} for $\Cc_j$ with noise-weighting
by measurement covariance $\V_j$
is a straight-forward
linear least-squares problem which may be solved with Singular
Value Decomposition (SVD), QR factorization, conjugate gradients, or
other methods.  For example, using the method of
``Normal equations''\footnote{Note that this method is mathematically
correct but numerically unstable.  It is included here for illustration,
but the actual calculation should use one of the other methods \citep{NRC2002}.}
and the shorthand $\vec x = \Xc_j$ and $\vec c = \Cc_j$:
\begin{eqnarray}
    \vec x & = & \PP \vec c \\
    \V^{-1} \vec x & = & \V^{-1} \PP \vec c \\
    \PP^T \V^{-1} \vec x & = & (\PP^T \V^{-1} \PP) \vec c \\
    (\PP^T \V^{-1} \PP)^{-1} \PP^T \V^{-1} \vec x & = & \vec c \\
\end{eqnarray}

If the noise is independent between variables, the
inverse covariance $\V^{-1}$ is just a diagonal matrix of weights $\Wc_j$.
Note that the covariance here is the estimated \emph{measurement}
covariance, not the total dataset variance --- the goal is to weight
the observations by the estimated \emph{measurement} variance so that noisy
observations do not unduly affect the solution, while allowing PCA to
describe the remaining \emph{signal} variance.

In the more general case of measurement covariance between different
observations, one cannot solve equation \ref{eq:solve_c} for each
column of $\X$ independently.  Instead, solve $\vec \X = [\PP] \vec \C$ with
the full covariance matrix $\mathbf{V}$ of $\vec \X$,
where $[\PP]$ is the matrix
formed by stacking $\PP$ $\nobs$ times, and $\vec\X$ and $\vec\C$ are
the vectors formed by stacking all the columns of the matrices $\X$ and $\C$.
This requires the solution of a single
$(\nobs \cdot \nvec) \times (\nobs \cdot \nvec)$ matrix rather than
$\nobs$ solutions of $\nvec \times \nvec$ matrices.  If the individual
observations are uncorrelated, it is computationally advantageous to use
this non-correlation to solve multiple smaller matrices
rather than one large one.

\subsubsection{Notes on solving $\{\vec \phi_k\} = \PP$}
\label{sec:solve_p}

\begin{figure}[t]
\centering
\plotone{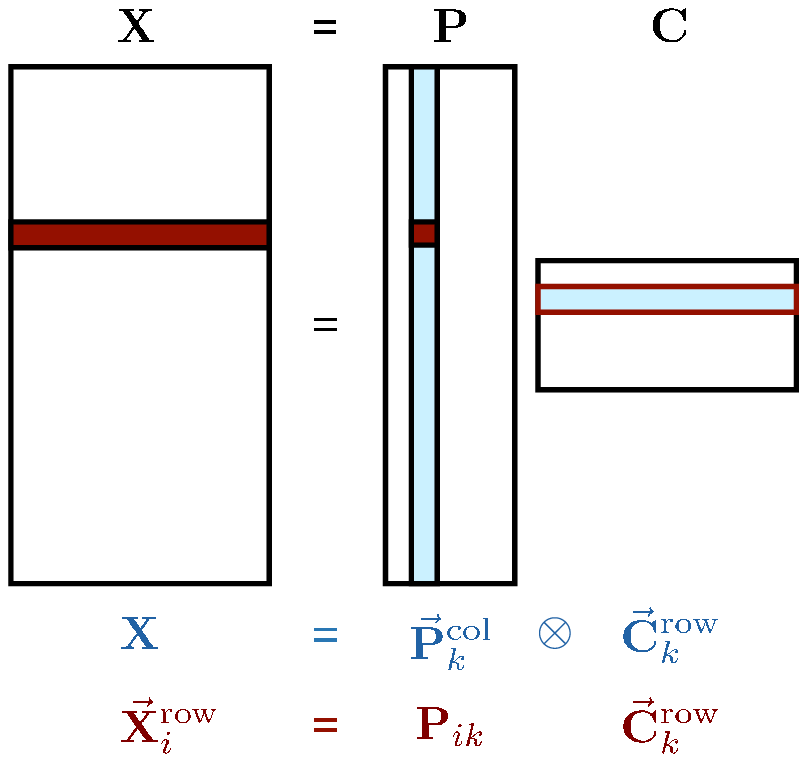}
\caption{Illustration of solving $\PP$ one element at a time,
{\it i.e.,} equation \ref{eq:solve_pik}.}
\label{fig:solve_p}
\end{figure}

In the second step of each iteration (equation \ref{eq:maximization}),
we use the fixed coefficients
$\C$ (dimensions $\nvec \times \nobs$) and solve for
the eigenvectors $\PP$ (dimensions $\nvar \times \nvec$).
We solve the eigenvectors one-by-one
to maximize the power in each eigenvector before solving the next.
Selecting out just the $k$th eigenvector uses the $k$th column of
$\PP$ and the $k$th row of $\C$:
\begin{eqnarray}
    \X & = & \PPc_k \otimes \Cr_k            \label{eq:solve_p}
\end{eqnarray}
where $\otimes$ signifies an outer product.
If the variables (rows) of $\X$ are independent, then we can solve for
a single element of $\PPc_k$ at a time:
\begin{eqnarray}
    \label{eq:solve_pik}
    \Xr_i & = & \PP_{ik} \Cr_k
\end{eqnarray}
This is illustrated in figure~\ref{fig:solve_p}.
With independent weights $\Wr_i$ on the data $\Xr_i$, we solve
variable $i$ of eigenvector $k$ with:
\begin{equation}
    \label{eq:solve_weighted_pik}
    \PP_{ik} =
    { \sum_j \W_{ij} \X_{ij} \C_{ik} \over
      \sum_j \W_{ij} \C_{ik} \C_{ik} }
\end{equation}

As with section \ref{sec:solve_c}, if there are measurement
covariances between the data, equation \ref{eq:solve_p} may be
expanded to solve for all elements of $\PPc_k$ simultaneously using
the full measurement covariance matrix of $\X$.

After solving for $\PPc_k$, subtract its projection from the data:
\begin{equation}
    \X \leftarrow \X - \PPc_k \otimes \Cr_k
\end{equation}
This removes any variation of the data in the direction of
$\PPc_k$ so that additional eigenvectors will be orthogonal to
the prior ones.\footnote{Note that this approach is potentially susceptible
to build up of machine rounding errors and should be explicitly
checked when using EMPCA for solving a large number of eigenvectors.}
Then repeat the procedure to solve for the next eigenvector $k+1$.

\section{Extensions of Weighted EMPCA}

The flexibility of the iterative EMPCA solution allows for a number of
powerful extensions to PCA in addition to noise weighting.  We describe
a few of these here.

\subsection{Smoothed EMPCA}
\label{sec:smooth}

If the length scale of the underlying signal eigenvectors
is larger than that of the noise, it may be advantageous to
smooth the eigenvectors to remove remaining noise effects.
The iterative nature of EMPCA allows
smoothing of the eigenvectors at each step to remove the high frequency
noise.  This generates the optimal smooth eigenvectors by construction
rather than smoothing noisy eigenvectors afterward.
This will be shown in the examples in section \ref{sec:examples}.
Alternately, one can include a smoothing prior or regularization term
when solving for the principal components $\PP$.  That approach,
however, requires solving equation \ref{eq:solve_p}
(plus a regularization term)
for all elements of
$\PPc_k$ simultaneously instead of using the numerically much faster
equation \ref{eq:solve_weighted_pik}
for the case of diagonal measurement covariance.

\subsection{A priori eigenvectors}
\label{sec:apriori}

In some applications, one has a few {\it a priori} template vectors to
include in the fit, \eg, from some physical model.  The goal is to find
additional template vectors which are to be combined with the {\it a priori}
vectors for the best fit of the data.  Due to noise weighting and the potential
non-orthogonality of the {\it a priori} vectors, the best fit is a joint
fit and one cannot simply fit the {\it a priori} vectors and remove their
components before proceeding with finding the other unknown vectors.

This case can be incorporated into EMPCA by including the {\it a priori}
vectors in the starting vectors $\PP$ and simply keeping them fixed with
each iteration rather than updating them.  In each iteration, the
\emph{coefficients} for the {\it a priori} vectors are updated, but not the 
vectors themselves.

%

\section{Examples}
\label{sec:examples}

\subsection{Toy data}

\begin{figure}[t]
\centering
\plotone{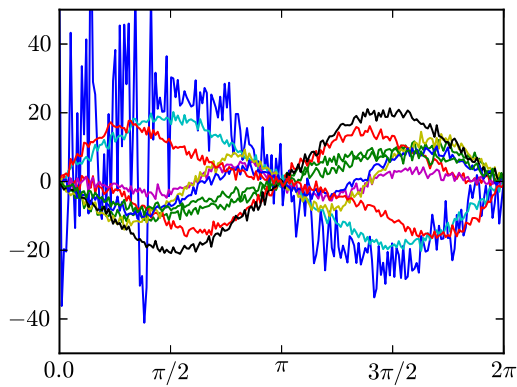}
\caption{Example of noisy data used to test weighted EMPCA.}
\label{fig:toy_data}
\end{figure}

\begin{figure}[t]
\centering
\plotone{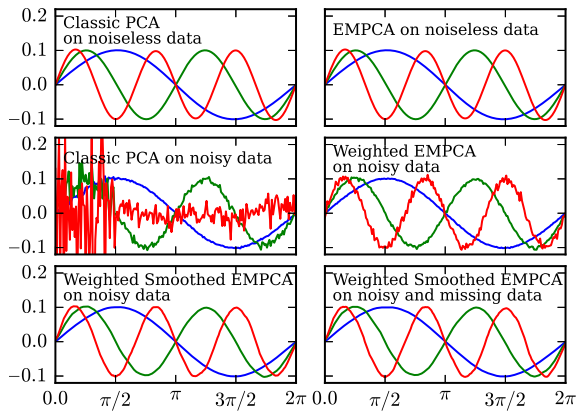}
\caption{Examples of classic PCA and EMPCA applied to noiseless, noisy,
and missing data.}
\label{fig:toy_pca}
\end{figure}

Figure \ref{fig:toy_data} shows example noisy data used to test weighted EMPCA.
100 data vectors were generated
using 3 orthonormal sine functions as input, with random amplitudes drawn
from Gaussian distributions.  The lower frequency sine waves were given
larger Gaussian sigmas such that they contribute more signal variance to
the data.  Gaussian random noise was added, with 10\% of
the data vectors receiving 25 times more noise from [0, $\pi/2$] and
5 times more noise from [$\pi/2$, $2\pi$].
For Weighted EMPCA, weights were assigned as $1/\sigma^2$, where
$\sigma$ is the per-observation per-variable Gaussian sigma of the
added noise (not the sigma of the underlying signal).
A final dataset was created where a contiguous 10\% of
each observation was set to have weight=0 to create
regions of missing data.  As a crosscheck that the 0 weights
are correctly applied, the data in these regions were set to a constant
value of 1000 -- if these data are not correctly ignored by the algorithm
they will have a large effect on the extracted eigenvectors. 

Figure \ref{fig:toy_pca} show the results of applying classic PCA
and weighted EMPCA to these data.
{\it Upper left}: Classic PCA applied to the noiseless data recovers
  the input eigenvectors, slightly rotated to form the best ranked
  eigenvectors for describing the data variance.
{\it Upper right}: EMPCA applied to the same noiseless data recovers
  the same input eigenvectors.
{\it Middle left}: When classic PCA is applied to the noisy data,
  the highest order eigenvector is dominated by the noise, and the
  effects of the non-uniform noise are clearly evident as increased
  noise from [0, $\pi/2$].
{\it Middle right}: Weighted EMPCA is much more robust to the noisy
  data, extracting results close to the original eigenvectors.  The
  highest order eigenvector is still affected by the noise, which is
  a reflection that the noise does contribute power to the data
  variance.  However, the extra-noisy region from [0, $\pi/2$] is
  not affected more than the region from [$\pi/2$, $2\pi$], due to
  the proper deweighting of the noisy data.
{\it Lower left}: Smoothed weighted EMPCA is almost completely
  effective at extracting the original eigenvectors with minimal
  impact from the noise.
{\it Lower right}: Even when 10\%
  of every observation is missing, smoothed weighted EMPCA
  is effective at extracting the underlying eigenvectors.
All eigenvectors for all methods are orthogonal at the level of
${\cal O}(10^{-17})$.

\subsection{QSO data}
\label{sec:qso}

\begin{figure}[t]
\centering
\plotone{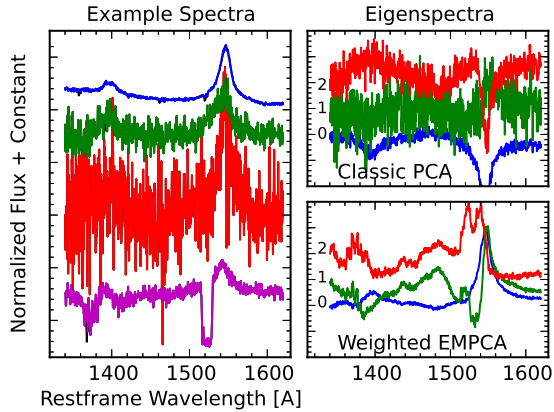}
\caption{Example of high-, median-, and low-$S/N$ input QSO spectra
and a broad absorption line (BAL) QSO (left).
The first 3 classic PCA eigenvectors are shown at top right;
the first 3 weighted EMPCA eigenvectors are shown at bottom right.
}
\label{fig:qso_pca}
\end{figure}

Figure~\ref{fig:qso_pca} shows the results of applying classic PCA
and weighted EMPCA to QSO spectra from the Sloan Digital Sky Survey
Data Release 7 \citep{DR7}, using the QSO redshift catalog of
\cite{HewettWild2010}.
500~spectra of QSOs with redshift
$2.0<z<2.1$ were randomly selected and trimmed to
$1340 < \lambda < 1620$~\AA~to show the SiIV and CIV
emission features.
Spectra with more than half of the pixels masked were discarded.
Each spectrum was normalized to
median[flux($1440 < \lambda < 1500$~\AA)]~$= 1$
and the weighted mean of all normalized spectra was subtracted.
The left panel of
Fig.~\ref{fig:qso_pca} plots examples of high, median, and low
signal-to-noise spectra and a
broad absorption line (BAL) QSO from this sample.
$\sim$2\% of the spectral bins have been flagged with a bad-data mask,
\eg, due to cosmic rays, poor sky subtraction, or the presence of
non-QSO narrow absorption features from the intergalactic medium.
These are treated as missing data with weight=0.
The goal of weighted EMPCA
is to properly deweight the noisy spectra such that the resulting
eigenvectors are predominantly describing the underlying signal variations
and not just measurement noise.
Weights are $1/\sigma_{ij}^2$ where $\sigma_{ij}$
is the SDSS pipeline estimated measurement noise for wavelength bin $i$
of spectrum $j$.
Weighted EMPCA can also properly ignore the masked data by using
weight=0 without having to discard the entire spectrum, or artificially
interpolate over the masked region.

The right panels of Fig.~\ref{fig:qso_pca} show the results for the
first 3 eigenvectors of classic PCA (top right) and weighted EMPCA
(bottom right).  Eigenvectors 0, 1, and 2 are plotted in blue, green,
and red respectively.  The mean spectrum was subtracted prior
to performing PCA such that these eigenvectors represent the principal
variations of the spectra with respect to that mean spectrum.
Eigenvectors are orthogonal at the level of ${\cal O}(10^{-17})$.

The EMPCA eigenvectors are much less noisy than the classic PCA
eigenvectors.  As such, they are more sensitive to genuine signal
variations in the data.  For example,
the sharp features between $1515 < \lambda < 1545$~\AA\ in
the EMPCA eigenspectra arise from
BAL QSOs, an example of which is shown
in the bottom of the left panel of Fig.~\ref{fig:qso_pca}.
These features are used to study QSO outflows,
\eg\ \citet{Turnshek1988}, yet they are
lost amidst the noise of the classic PCA eigenspectra.
Similarly, the EMPCA eigenspectra are  more sensitive to the details of the
variations in shape and location of the emission peaks
used to study QSO metallicity (\eg\ \citet{Juarez2009})
and black hole mass (\eg\ \citet{Vestergaard2009}).

\section{Discussion}
\label{sec:discussion}

\subsection{Convergence}
\label{sec:convergence}
\citet{MK97} discuss the convergence properties of the
EM algorithm in general.  Each iteration, by construction, finds
a set of parameters that are as good or better a fit to the data than
the previous step, thus guaranteeing convergence.  The caveat
is that the ``Likelihood maximization step'' is typically implemented
as solving for a stationary point of the likelihood surface rather
than strictly a maximum.
\eg, $\partial {\cal L} / \partial \phi = 0$ is also true at saddle
points and minima of the likelihood surface, thus it is possible that
the EM algorithm will not converge to the true global maximum.
Unweighted PCA has a likelihood surface with a single
global maximum, but in general this is not the case for weighted PCA:
the weights in equation~\ref{eq:weighted_chi2} can result in local false
$\chi^2$ minima \citep{SrebroJaakkola2003}.
\citet{MK97} \S 3.6 also gives examples of this behavior
(taken from \citet{Murray1977} and \citet{ACK1993})
for the closely related problem of Factor Analysis.
The example datasets are somewhat contrived and the minimum or saddle point
convergence only happens with particular starting conditions.

We have encountered false minima with weighted EMPCA
when certain observations have $\sim$90\% of their variables masked
while giving large weight to their remaining unmasked variables.
In this case the resultant eigenvectors can have artifacts tuned
to the highly weighted but mostly masked input observations.
When only a few ($\sim$10\%) of the variables are masked per observation,
we have not had a problem with false minima.

The algorithm outlined in section \ref{sec:WEMPCA}
solves for each eigenvector one at a time in order to maximize the power
in the initial eigenvectors.  This can result in a situation where a
given iteration can improve the power described by the first few eigenvectors
while degrading the total $\chi^2$ using all eigenvectors.  We have not
found a case where this significantly degrades the global $\chi^2$, however.

The speed of convergence is also not guaranteed.  \citet{Roweis1997} gives a
toy example of fast convergence for Gaussian-distributed data (3 iterations),
and slow convergence for non-Gaussian-distributed data (23 iterations).
In practice we find that when EMPCA is slow to converge, it is exploring a
shallow likelihood surface between two nearly degenerate eigenvectors.
This situation pertains to the uniqueness of the solution, described in
the following section.

Weighted EMPCA may produce unstable solutions if it is used to solve for more
eigenvectors than are actually present in the data, or for eigenvectors
that are nearly singular.  Since EMPCA uses all eigenvectors while solving
for the coefficients during each iteration, the singular eigenvectors can
lead to delicately balanced meaningless values of the coefficients, which
in turn degrades the solution of the updated eigenvectors in the next
iteration.  We recommend starting with solving for a small number of
eigenvectors, and then increasing the number of eigenvectors if the resulting
solution does not describe enough of the data variance.

For these reasons, one should use caution when analyzing data with EMPCA,
just as one should do with any problem which is susceptible to false minima
or other convergence issues.  In practice, we find that the benefits of
proper noise-weighting outweigh the potential convergence problems.

\subsection{Uniqueness}
\label{sec:uniqueness}
Given that EMPCA is an iterative algorithm with a random starting point,
the solution is not unique.  In particular, if two eigenvalues are very
close in magnitude, EMPCA could return an admixture of the
corresponding eigenvectors while still satisfying the convergence criteria.
In practice, however, EMPCA is pragmatic:
if two eigenvectors have the same eigenvalue, they are also equivalently
good at describing the variance in the data and could be used interchangeably.

Science applications, however, generally require strict algorithmic reproducibility
and thus EMPCA should be used with a fixed random number generator seed
or fixed orthonormal starting vectors such as Legendre polynomials.
The convergence criteria define when a given
vector is ``good enough'' to move on to the next iteration, but they do
not guarantee uniqueness of that vector.

\begin{figure}[t]
\centering
\plotone{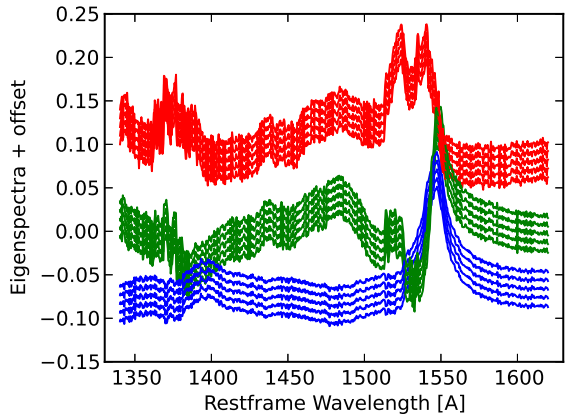}
\caption{Example eigenspectra of the same dataset with different
random starting values for the EMPCA algorithm.  The eigenspectra
are offset for clarity and have bin-for-bin agreement at the level
of $\sim$10$^{-5}$.
}
\label{fig:qso_convergence}
\end{figure}

Figure \ref{fig:qso_convergence} shows the first 3 eigenvectors for
5 different EMPCA solutions of the QSO spectra from section \ref{sec:qso}
with different random starting vectors.
After 20 iterations the eigenvectors agree to
$<10^{-5}$ on both large and small scales.
Although this agreement is worse than the machine precision of
the computation, it is much smaller than the scale of differences
between the eigenvectors and it represents a practical level of
convergence for most PCA applications.

\subsection{Over-weighted Data}
\label{sec:overweighting}

Weighted EMPCA improves the PCA eigenvector solution by
preventing noisy or missing data from unduly contributing noise instead of
signal variation.
However, the opposite case of high signal-to-noise data can also be
problematic if just a few of the observations have significantly higher
weight than the others.
These will dominate the
EMPCA solution just as they would dominate a weighted mean calculation.
This may not be the desired effect since the initial eigenvectors will
describe the differences between the highly weighted data and
subsequent eigenvectors will describe the lower weighted data.
This may be prevented by purposefully down-weighting certain observations
or applying an upper limit to weights so that the weighted dataset isn't
dominated by just a few observations.

\subsection{Scaling Performance}
\label{sec:scaling}

The primary advantage of EMPCA is the ability to incorporate weights
on noisy data to improve the quality of the resulting eigenspectra.
A secondary benefit over classic PCA
is algorithmic speed for the common case of needing
only the first few eigenvectors from a dataset with
$\nobs \stackrel{<}{\sim} \nvar$.

Classic PCA requires solving the eigenvectors of the data covariance matrix,
an ${\cal O}(\nvar^3)$ operation.
The weighted EMPCA algorithm described here involves iterating over multiple
solutions of smaller matrices.  Each iteration requires
$\nobs$ solutions of ${\cal O}(\nvec^3)$ to solve the coefficients and
${\cal O}(\nobs \nvec \nvar)$ operations to solve the eigenvectors.
Thus weighted EMPCA can be faster than classic PCA when
$n_\mathrm{iter} (\nobs \nvec^3 + \nobs \nvec \nvar) < \nvar^3$, ignoring the constant prefactors.
If one has a few hundred spectra ($\nobs$)
with a few thousand wavelengths each ($\nvar$)
and wishes to solve for the first few eigenvectors ($\nvec$),
then EMPCA can be much faster than classic PCA.
Conversely, if one wishes to perform PCA on all $\sim$1 million spectra from SDSS,
then $\nobs \gg \nvar$ and classic PCA is faster, albeit with the limitations
of not being able to properly weight noisy or missing data.
If the problem involves off-diagonal covariances, then weighted EMPCA
involves a smaller number of larger matrix solutions for an overall
slowdown, though note that classic PCA is unable to properly solve the
problem at all.

As a performance example, we used EMPCA to study the variations in the
simulated point spread function (PSF) of a new spectrograph design.
The PSFs were simulated on a grid of 11 wavelengths and 6 slit locations,
and were sampled over $200 \times 200$~$\mu$m spots
on a 1~$\mu$m grid, for a total of 40000 variables per spot.
Classic PCA would require singular value decomposition of a
$40000 \times 40000$ matrix.  While this is possible, it is
beyond the scope of a typical laptop computer.  On the other hand, using
EMPCA with constant weights we were able to recover the first
30 eigenvectors covering 99.7\% of the PSF variance in less than
6 minutes on a 2.13 GHz MacBook Air laptop.

For datasets where $\nvar$ is particularly large, the memory needed to store
the $\nvar \times \nvar$ covariance matrix may be a limiting factor for
classic PCA.  The iterative nature of EMPCA allows one to scale to extremely
large datasets since one never needs to keep the entire dataset
(nor its covariance) in memory at one time.  The multiple independent
equations to solve in sections \ref{sec:solve_c} and \ref{sec:solve_p}
are naturally computationally parallelizable.

\section{Python Code}

Python code implementing the Weighted EMPCA algorithm described here
is available at
\url{https://github.com/sbailey/empca}~.
The current version implements the case of independent
weights but not the more generalized case of off-diagonal covariances.
It also implements the smoothed eigenvectors described in \S\ref{sec:smooth},
but not {\it a priori} eigenvectors (\S\ref{sec:apriori}),
nor distributed calculations (\S\ref{sec:scaling}).
For comparison, the {\tt empca} module also includes implementations
of classic PCA and weighted lower-rank matrix approximation.
Examples for this paper were prepared with tagged version {\tt v0.2}
of the code.

When using the code, note that the orientation of the data and weights
vectors is the transpose of the notation used here, {\it i.e.},
{\tt data[j,i]} is variable {\tt i} of observation {\tt j} so that
{\tt data[j]} is a single observation.

\section{Summary}

To briefly summarize the algorithm:  A data matrix $\X$
can be approximated by a set of eigenvectors $\PP$ with coefficients $\C$:
\begin{equation}
    \X \approx \PP \C + \mathrm{measurement~noise}
\end{equation}
A covariance matrix $\V$ describes the estimated measurement noise.

The Weighted EMPCA algorithm seeks to find the optimal $\PP$ and $\C$
given $\X$ and $\V$.
It starts with a random set of orthonormal
vectors $\PP$, and then iteratively alternates solutions for
$\C$ given $\{\PP, \X, \V\}$
and $\PP$ given $\{ \C, \X, \V \}$.
The problem is additionally constrained by the requirement to
maximize the power in the fewest number of eigenvectors (columns of $\PP$).
To accomplish this, the algorithm
solves for each eigenvector individually, before
removing its projection from the data and solving for the next eigenvector.
If the measurement errors are independent, the covariance can be
described by a weights matrix $\W$ with the same dimensions as $\X$,
and the problem can be factorized into independent solutions of small
matrices.

This algorithm produces a set of orthogonal principal component eigenvectors $\PP$, which are optimized to describe the most signal variance with the
fewest vectors while properly accounting for estimated measurement noise. 

\section{Conclusions}

We have described a method for performing PCA on noisy data that properly
incorporates measurement noise estimates when solving for the eigenvectors
and coefficients.  Missing data is simply the limiting case of weight=0.
The method uses an iterative solution based upon Expectation Maximization.
The resulting eigenvectors are less sensitive to measurement noise and
more sensitive to true underlying signal variations.  The algorithm has
been demonstrated on toy data and QSO spectra from SDSS.
Code which implements this algorithm is available at
\url{https://github.com/sbailey/empca}~.

\section {Acknowledgements}

The author would like to thank Rollin Thomas and S\'{e}bastien Bongard for
interesting and helpful conversations related to this work.  The
anonymous reviewer provided helpful comments and suggestions which
improved this manuscript.  The initial
algorithm was developed during a workshop at the Institut de Fragny.
This work was supported under the auspices of the Office of Science,
U.S. DOE, under Contract No. DE-AC02-05CH1123.

The example QSO spectra were provided by the Sloan Digital Sky Survey.
Funding for the SDSS and SDSS-II has been provided by the Alfred P. Sloan
Foundation, the Participating Institutions, the National Science Foundation, the
U.S. Department of Energy, the National Aeronautics and Space Administration,
the Japanese Monbukagakusho, the Max Planck Society, and the Higher Education
Funding Council for England. The SDSS Web Site is http://www.sdss.org/.

The SDSS is managed by the Astrophysical Research Consortium for the
Participating Institutions. The Participating Institutions are the American
Museum of Natural History, Astrophysical Institute Potsdam, University of Basel,
University of Cambridge, Case Western Reserve University, University of Chicago,
Drexel University, Fermilab, the Institute for Advanced Study, the Japan
Participation Group, Johns Hopkins University, the Joint Institute for Nuclear
Astrophysics, the Kavli Institute for Particle Astrophysics and Cosmology, the
Korean Scientist Group, the Chinese Academy of Sciences (LAMOST), Los Alamos
National Laboratory, the Max-Planck-Institute for Astronomy (MPIA), the
Max-Planck-Institute for Astrophysics (MPA), New Mexico State University, Ohio
State University, University of Pittsburgh, University of Portsmouth, Princeton
University, the United States Naval Observatory, and the University of
Washington.


\end{document}